\pdfoutput=1
\documentclass[11pt,aps,a4paper,eqsecnum,amsmath,amssymb
,superscriptaddress,notitlepage,nofootinbib,longbibliography
]{revtex4-1}

\usepackage{graphicx}

\begin{document}
\preprint{}
\title{$OSp(n|2m)$ quantum chains with free boundaries}  

\author{Holger Frahm}
\affiliation{%
Institut f\"ur Theoretische Physik, Leibniz Universit\"at Hannover,
Appelstra\ss{}e 2, 30167 Hannover, Germany}

\author{M\'arcio J. Martins}
\affiliation{%
Departamento de F\'isica, Universidade Federal de S\~ao Carlos,
C.P. 676, 13565-905 S\~ao Carlos (SP), Brazil}


\begin{abstract}
In this paper we investigate the spectrum of $OSp(n|2m)$ quantum spin chains with free boundary conditions. We compute the surface free energy of these models which, similar to other properties in the thermodynamic limit including the effective central charge of the underlying conformal field theory, depends on $n-2m$ only. For several models in the regime $n-2m< 2$ we have studied the finite-size properties including the subleading logarithmic corrections to scaling. As in the case of periodic boundary conditions we find the existence of a tower of states with the same conformal dimension as the identity operator. As expected the amplitudes of the corresponding logarithmic corrections differ from those found previously for the models with periodic boundary conditions. We point out however the existence of simple relations connecting such amplitudes for free and periodic boundaries.
  Based on our findings we formulate a conjecture on the long distance behaviour of the bulk and surface watermelon correlators.
\end{abstract}


\maketitle

\section{Introduction}
In recent years there has been some interest in the study of the
critical properties of integrable one-dimensional quantum spin chains
based on supergroup symmetries because of their mathematical and
physical implications.  For instance, the staggered $sl(2|1)$
superspin chain with spins alternating between the fundamental and
dual representations may be of relevance for the description of
properties of fermions in the presence of random potentials
\cite{Gade99,EsFS05}. Yet another example are the spin chains
invariant by the fundamental vector representation of the $OSp(n|2m)$
superalgebra which can be related to an intersecting loop model on the
square lattice with fugacity $z=n-2m$ \cite{MaNR98}. This loop model
describes the motion of particles through randomly fixed scatterers in
such way that path intersections are allowed. For $n-2m<2$ the
crossing of the loops appears to become a relevant perturbation and
model properties have been argued to be those of the Goldstone phase
of the $O(z)$ sigma model \cite{JaRS03}.  The spectrum of these
superspin chains for large number of sites $L$ present some
distinguished features when compared to that of spin chains based on
ordinary Lie groups. For example, it was observed that several of the
scaled gaps appear to produce the same conformal weight implying a macroscopic degeneracy of the respective state
in the thermodynamic limit \cite{MaNR98,FrMa18}.  A similar observation in the staggered $sl(2|1)$ superspin chain has been interpreted as signature for a continuous component to the spectrum of conformal weight resulting from a non-compact target space of the
conformal field theory associated to these superspin chains \cite{EsFS05,SaSc07}.
At this point we remark that such scenario has also been found in
other families of staggered vertex models
\cite{IkJS08,FrMa11,FrMa12,FrHo17} and quantum deformations of
superspin chains \cite{VeJS14,VeJS14a,VeJS16a,FrHM19}.

We further motivate this work by mentioning some of our earlier findings
concerning the eigenspectrum behaviour of $OSp(n|2m)$ superspin chains
with periodic boundary conditions \cite{FrMa15,FrMa18}: in these
models there exist towers of low energy excitations over the ground
state which for large system sizes leads to the same effective central
charge $c_{\mathrm{eff}}$. If we denote the energies of such set of
states by $E_{k}(L)$ we have found that the behaviour for
${L} \rightarrow \infty$ is
\begin{equation}
\label{Elowest-CFT}
 E_{k}(L) = L e_{\infty} + \frac{2 \pi v_F}{L}\left(-\frac{c_{\mathrm{eff}}}{12}
  +\frac{\beta(k)}{\log L} \right) \,,
  \quad k=0,1,2,\cdots,k_{\infty}
\end{equation}
where the integer $k_{\infty}$ is limited by system size $\mathrm{L}$,
$e_{\infty}$ refers to the ground state energy per site in the
thermodynamic limit and $v_F$ denotes the velocity of the low-lying
excitations.
We next observe that in the regime $n-2m<2$ certain correlation
functions of the related loop model can be rewritten in terms of the
subleading logarithmic amplitudes $\beta_k$. These 'watermelon
correlators' measure the probability of $k$ distinct loop segments
connecting two arbitrary lattice points $x$ and $y$ which for large
distances $r=|x-y|$ which has been argued to decrease logarithmically
with $r$ \cite{Polyakov75,NSSO13}.  As we have pointed out in
Ref.~\cite{FrMa18} this behaviour of bulk correlation functions can be re-written in terms of the
finite-size logarithmic amplitudes as follows
\begin{equation}
G^{(b)}_{k}(r) \sim 1/\ln(r)^{2(\beta(k)-\beta(k_0))}\,
\end{equation}
for a suitable choice of the $k_0$ state. 

The purpose of the present paper is to investigate the effect of boundary conditions on the spectrum of conformal weights of the $OSp(n|2m)$ superspin chains. Generally, knowing the properties of a critical system under various boundary conditions is a prerequisite for the identification of the full operator content of a given universality class \cite{Card84,Card86b}. Moreover, recent studies of the staggered six-vertex model have revealed that open boundary conditions may change its low energy properties significantly \cite{RPJS20,RoJS21,FrGe22}.  Here we shall present evidence that the tower of low energy states over the identity operator present in the $OSp(n|2m)$ models with periodic boundary conditions continues to exist in the presence of free boundary conditions.  
As we shall argue these states have the following finite-size structure as $L\to\infty$:
\begin{equation}
\label{EFRE}
 E_{k}(L) = L e_{\infty} + f_{\infty} +\frac{\pi v_F}{L}\left(-\frac{c_{\mathrm{eff}}}{24}
  +\frac{\alpha(k)}{\log L} \right) \,,
  \quad k=0,1,2,\cdots,k_{\infty}
\end{equation}
where $f_{\infty}$ is the surface energy resulting from the free
boundary conditions.
%
The leading finite-size term in (\ref{EFRE}) is in
accordance with the predictions for conformally invariant theories
with free boundary conditions \cite{BlCN86}.  Even for conventional conformal theories, however, the subleading corrections are expected to depend on the boundary terms \cite{Card86c,AfQi99}.
Indeed, we find that the amplitudes $\alpha_k$
and $\beta_k$ differ.  For $n-2m<2$ we present evidences
that they appear to obey the rather simple relation,
\begin{equation}
\alpha(k)-\alpha({k_0}) =
2(\beta(k)-\beta({k_0}))\,,\quad n-2m<2
\end{equation}
for the similar choice of the state $k_0$ for both 
free and periodic 
boundary conditions.

It is now tempting to use the above relationship 
among logarithmic 
amplitudes and
the asymptotic behaviour of correlators to infer about
the behaviour of surface watermelon correlators 
for large distances. Recall here
that free boundary conditions 
play the role of Dirichlet boundary conditions in which the 
order parameters entering the correlators are expected
to vanish on the boundary.
Let us denote the surface watermelon correlator by
$G^{(s)}_{k}(\rho)$ where $\rho$ is the distance 
between to points $x$ and $y$ 
parallel the half-plane boundary. Considering that 
the asymptotic behaviour 
of such correlators should be governed instead
by the surfaces amplitudes $\alpha_k$ one obtains, 
\begin{equation}
G^{(s)}_{k}(\rho) \sim 1/\ln(\rho)^{4(\beta({k})-\beta({k_0}))}\,
\end{equation}
and hence a faster logarithmic surface decay as compared with the bulk
behaviour by a factor two.  Note that this dependence on $\rho$ is very different from that of polymers, i.e.\ loops without intersections \cite{DuSa86}.

\section{The open $OSp(n|2m)$ spin chain properties }

In this section we describe the thermodynamic limit 
properties of  
spin chains based on the vector representation of the 
$OSp(n|2m)$ superalgebra 
with free boundary conditions. The model Hamiltonian
can be represented in terms of generators of 
a braid-monoid algebra
which underpins a square lattice loop model
admitting intersections between the polygon
configurations \cite{MaNR98}. 
The Hamiltonian of the spin chain in an one-dimensional lattice of size $L$ is given by
\begin{equation}
\label{HAM}
\mathrm{H}= \epsilon \sum_{i=1}^{L-1} \left[ \mathrm{P}_{i,i+1}+\frac{2}{2-z}\mathrm{E}_{i,i+1} \right]\,,
\end{equation}
where we chose $\epsilon$ to select the anti-ferromagnetic regime of the model, i.e.\ $\epsilon=-1$ $(+1)$ for $n-2m<2$ ($>2$).
Note that Eq.~(\ref{HAM}) describes the superspin 
chain with free 
boundary 
conditions. 
The fugacity $z$ of the related intersecting loop model is
realized in the spin chain as the difference between the number of the
bosonic and fermionic degrees of freedom $z=n-2m$. 

The braid $\mathrm{P}_{i,i+1}$ turns out to the graded permutation operator whose expression is,
\begin{equation}
\mathrm{P}_{i,i+1}= \sum_{\alpha,\beta=1}^{n+2m} (-1)^{p_{\alpha} p_{\beta}} 
e_{\alpha \beta} \otimes 
e_{\beta \alpha}
\end{equation}
where $p_{\alpha}$ are the Grassmann parities for 
the $n$ bosonic ($p_{\alpha}=0$) 
and the $2m$ fermionic ($p_{\alpha}=1$) degrees of freedom. 
The matrices $e_{\alpha \beta }$ have only one non-vanishing
element with value 1 at row $\alpha$ and
column $\beta$.
The operator $\mathrm{E}_{i,i+1}$ is a generator 
of the Temperley-Lieb algebra weighted by the fugacity $z$. It can be
represented by the expression,
\begin{equation}
\mathrm{E}_{i,i+1}= \sum_{\alpha,\beta,\gamma,\delta=1}^{n+2m} 
A_{\alpha \beta} A^{-1}_{\gamma \delta}e_{\alpha \gamma} \otimes 
e_{\beta \delta}
\end{equation}
where the non-zero matrix elements $A_{\alpha \beta}$ 
are $\pm 1$ such that their matrix positions depend on the 
grading ordering
of the basis. For explicit matrix representations of the Temperley-Lieb 
generator see for
instance \cite{MaRa97a}.

Before proceeding we remark that the quantum integrability of the
Hamiltonian (\ref{HAM}) can be established within the double row
transfer matrix framework devised by Sklyanin for the Heisenberg chain
\cite{Skly88}.  In this method the Hamiltonian boundary terms depend
on the certain one-body scattering matrices on the half-line. In the
specific case of free boundary conditions considered in this paper
these reflecting matrices are trivial being proportional to the
identity operators. For the details about the technical points
concerning this construction for the open $OSp(n|2m)$ spin chain see
for instance \cite{AACDF03,AACDF04}.

We have studied the eigenspectrum properties of open $OSp(n|2m)$ spin
chain (\ref{HAM}) for some values of the numbers $n$ of bosonic and $2m$ of fermionic
degrees of freedom.  Our numerical results for small lattice sizes
suggest we have the following sequence of spectral inclusions,
\begin{equation}
  \label{specinclusion}
\mathrm{Spec}[OSp(n|2m)] \subset \mathrm{Spec}[OSp(n+2|2(m+1))]  
\subset \mathrm{Spec}[OSp(n+4|2(m+2))] \subset \dots
\end{equation}
similar to what happens for periodic conditions \cite{FrMa18,GrJS19}.
As a consequence the basic properties of the Hamiltonian (\ref{HAM})
are expected to depend solely on the fugacity $z=n-2m$ in the
thermodynamic limit. In addition to that it is known that the
low-lying excitations of the $OSp(n|2m)$ with periodic boundary
conditions are gapless \cite{MaNR98,FrMa18}. This feature is not
expected to depend on the boundary conditions. As a consequence of
that the ground state energy of the Hamiltonian (\ref{HAM}) should
scale with the lattice size $L$ as \cite{BlCN86},
\begin{equation}
  E_0(L) \simeq L e_{\infty}+f_{\infty} -\frac{\pi v_F c_{\mathrm{eff}} }{24 L} \,,
\end{equation}
where $c_{\mathrm{eff}} $ is the effective central charge of the
respective conformal field theory. This invariant is expected to be
the same as the one underlying the model with periodic boundary
conditions \cite{MaNR98,FrMa18}
\begin{equation}
  \label{ceff-z}
  c_{\mathrm{eff}}=\begin{cases} 
    z/2 & \mathrm{for}~~~z \geq 2 \\
    z-1 & \mathrm{for}~~~z<2 \\
  \end{cases}\,.
\end{equation}

The parameters $e_{\infty}$ and $v_F$ denote the bulk ground state
energy and the Fermi velocity of the elementary excitations. Again, these bulk quantities are
expected not to depend of the boundary conditions, hence their
values are known to be \cite{MaNR98,FrMa18} given by
\begin{equation}
\label{einf}
e_{\infty}= -\frac{2}{|2-z|} \left[ \psi \left(\frac{1}{2} +\frac{1}{|2-z|} \right)
-\psi \left(\frac{1}{|2-z|} \right)
 +2\ln(2) \right]+1\,,
\end{equation}
where $\psi(x)$ is the Euler $\mathrm{psi}$ function, and the speed of
sound is $v_F=2\pi/|2-z|$.

By way of contrast the surface energy $f_{\infty}$ depends on the
boundary conditions which are imposed on the spin chain. Using the root density method \cite{YaYa69} we compute this quantity for the $OSp(1|2m)$, $OSp(2|2m)$ and various $O(n)$ models in Appendix~\ref{app:thermo}.  Together with our numerical results for the the $OSp(3|2)$ model this leads us to conjecture the expression of the surface energy for the generic
$OSp(n|2m)$ superspin chain with free boundary
conditions to be
\begin{equation}
\label{finf}
\begin{aligned}
f_{\infty} = -\frac{1}{2-z} 
&\left[ \psi \left(1 +\frac{1}{2(2-z)} \right)
-\psi \left(\frac{1}{2} +\frac{1}{2(2-z)} \right)
-\psi \left(\frac{3}{4}+\frac{1}{2(2-z)} \right) 
\right.  \\ 
&\left. 
+\psi \left(\frac{1}{4}+\frac{1}{2(2-z)} \right)
+2\ln(2) -\pi \right]+1,\qquad\mathrm{for~}z<2
\end{aligned}
\end{equation}
and
\begin{equation}
\label{finf2}
\begin{aligned}
f_{\infty} = -\frac{1}{z-2} 
&\left[ \psi \left(1 +\frac{1}{2(z-2)} \right)
-\psi \left(\frac{1}{2} +\frac{1}{2(z-2)} \right)
+\psi \left(\frac{3}{4}+\frac{1}{2(z-2)} \right) 
\right. \\ 
&\left . 
-\psi \left(\frac{1}{4}+\frac{1}{2(z-2)} \right)
+2\ln(2) -\pi \right]+1,\qquad\mathrm{for~}z>2
\end{aligned}
\end{equation}
Note the difference in signs of the last two Euler psi functions between the regimes $z<2$ and $z>2$: among the non-universal quantities describing the thermodynamics of the models the bulk energy and Fermi velocity of the $OSp(2|2(m-1))$ and $O(2m)$ spin chains coincide while their surface energies differ. The same is true for the effective central charge (\ref{ceff-z}) characteristic for the universal critical behaviour described by the underlying conformal field theory.

The model with $z=2$ has to be dealt with separately since the Hamiltonian (\ref{HAM}) becomes dominated by the Temperley-Lieb operator.
The simplest realization of this model is that with $n=2$ and $m=0$ 
which corresponds to the isotropic spin-$1/2$ Heisenberg model,
\begin{equation}
\label{XXX}
\mathrm{H}= \frac{1}{2}\sum_{i=1}^{L-1} \left[ \sigma_{i}^{x} \sigma_{i+1}^{x} 
+\sigma_{i}^{y} \sigma_{i+1}^{y} 
+\sigma_{i}^{z} \sigma_{i+1}^{z} -\mathrm{I}_{i,i+1} 
\right]
\end{equation}
where $\sigma_i^{x}, \sigma_i^{y}, \sigma_i^{z}$ are Pauli 
matrices acting
on the $i$-th lattice site and $\mathrm{I}_{i,i+1}$ is the $4 \times 4$ 
identity matrix.
It turns out that the bulk and the surface energies of this model may be obtained by considering the limit $z \rightarrow 2$ in Eqs.~(\ref{einf}) and (\ref{finf}). The coefficients proportional to $\mathcal{O}(1/\varepsilon)$ in the expansion around $z=2(1-\varepsilon)$ turns out to be the respective values for $e_{\infty}$ and $f_{\infty}$ associated to the Heisenberg chain (\ref{XXX}) \cite{Hult39,HaQB87}.  The Fermi velocity of massless excitations in this model is $v_F=\pi$.

In Table \ref{tab1} we present the parameters data characterizing the thermodynamic limit of some of the $OSp(n|2m)$ models including the ones whose critical properties we are going to analyze further below.
\begin{table}[t]
\begin{center}
\begin{tabular}{|c|c|c|c|c|}
  \hline
$z$ & $e_{\infty}$ & $f_{\infty}$ & $v_F$  & $c_{\mathrm{eff}}$  \\ \hline \hline
$-2$  & $-\frac{\pi}{2}-\ln(2) +1$  & $\frac{\pi}{4}(1+2\sqrt{2})-\frac{\ln(2)}{2}-1$ & $\frac{\pi}{2}$ & $-3$  \\ \hline
$-1$ & $-\frac{4\pi\sqrt{3}}{9}+1$ & $\pi+\frac{2\pi}{3\sqrt{3}}-\frac{2}{\sqrt{3}}\ln(2+\sqrt{3})-1$ & 
$\frac{2\pi}{3}$ & $-2$ \\ \hline 
$0$ & $-4\ln(2)$+1   &  $\pi-1$ & $\pi$ & $-1$\\ \hline
$1$ & $-3$   & $3$  & $2 \pi$ & $0$  \\ \hline
$2$ & $-2\ln(2)$  &  $\frac{\pi}{2}-\ln(2)$  & $\pi$ & $1$   \\ \hline
$3$ & $-3$  & $2\pi -5$  & $2 \pi $ &  $\frac{3}{2}$ \\ \hline
$4$ & $-4\ln(2)+1$  & $\pi-2\ln(2)-1$  & $\pi $ &  $2$ \\ \hline
$5$ & $-\frac{4\sqrt{3}}9\pi+1$  & $\frac{2 \pi}{\sqrt{3}}-\frac{\pi}{3}+\frac{2}{\sqrt{3}}\ln(2+\sqrt{3})-1$  & $\frac{2 \pi}{3} $ &  $\frac{5}{2}$ \\ \hline
\end{tabular}
\caption{The bulk and surface energies, the Fermi velocity as well as the effective central charge for some values of the fugacity.}
\label{tab1}
\end{center}
\end{table}

\section{Finite-size spectrum}
\label{sec:CFT}
We now turn to the analysis of the finite-size spectrum for the spin chains with fugacity $z<2$ exhibited in Table~\ref{tab1}.  The leading terms appearing in the finite-size scaling of low energy levels with quantum numbers $Q=\{q_1,q_2,\dots\}$ of a critical model in $1+1$ dimensions are given by conformal invariance \cite{BlCN86,Affl86,Card86a}: for periodic boundary conditions they are given as
\begin{equation}
  E_{Q}(L) \simeq L e_\infty + \frac{2\pi v_F}{L}
  \left( -\frac{c_{\mathrm{eff}}}{12} + X_{Q} + \dots
  \right)\,,
\end{equation}
while one has
\begin{equation}
\label{cft-open}
  E_{Q}(L) \simeq L e_\infty + f_\infty + \frac{\pi v_F}{L}
  \left( -\frac{c_{\mathrm{eff}}}{24} + X_{Q} + \dots\right)\,.
\end{equation}
in models with open boundary conditions.  Here $c_{\mathrm{eff}}$ is the effective central charge characterizing the universality class of the critical point and $X_Q$ are the (surface) critical dimensions describing the decay of correlations in the bulk and along the boundary, respectively.

For the $OSp(n|2m)$ spin chains the effective central charges are given by (\ref{ceff-z}) \cite{MaNR98,FrMa15,FrMa18}.  Therefore, the conformal weights (and possible subleading corrections to scaling) appearing in the models with free boundaries can be extracted from (\ref{cft-open}) by extrapolation of
\begin{equation}
    X_{\mathrm{eff},Q}(L) = \frac{L}{\pi v_F}\left(E_Q-Le_\infty-f_\infty\right) \equiv -\frac{c_{\mathrm{eff}}}{24} + X_Q(L)\,.
\end{equation}
Based on the perturbative RG analysis the model flows to weak coupling and our previous work on the periodic chains we expect logarithmic corrections to scaling, i.e.\
\begin{equation}
    X_Q(L) \simeq X_Q + \frac{\alpha(Q)}{\log L}+\dots\,,
\end{equation}
with integer conformal weights $X_Q$.  In the following we are particularly interested in the amplitudes $\alpha(Q)$ for the tower of levels over the identity operator with $X_Q=0$ and their relation to the corresponding ones found for the periodic spin chain \cite{FrMa18}.

%
\subsection{$z=-2$: the $OSp(2|4)$ superspin chain}
%
The Bethe equations for the $OSp(2|4)$ model in the grading $ffbbff$ read
\begin{equation}
  \label{bae24}
  \begin{aligned}
    &\left[f_{1/2}\left(\lambda_j^{(1)}\right)\right]^{2L} =
    \prod_{k\neq j}^{L-n_1} f_1\left(\lambda_j^{(1)}-\lambda_k^{(1)}\right)\,
    f_1\left(\lambda_j^{(1)}+\lambda_k^{(1)}\right)\times\\
    &\qquad \times
    \prod_{\sigma=\pm}\left(\prod_{k=1}^{L/2-n_\sigma}
    f_{-1/2}\left(\lambda_j^{(1)}-\lambda_k^{(\sigma)}\right)\,
    f_{-1/2}\left(\lambda_j^{(1)}+\lambda_k^{(\sigma)}\right)\right)\,,
    \quad j=1,\dots,L-n_1\,,\\
    &\prod_{k=1}^{L-n_1} f_{1/2}\left(\lambda_j^{(\pm)}-\lambda_k^{(1)}\right)\,
    f_{1/2}\left(\lambda_j^{(\pm)}+\lambda_k^{(1)}\right)
    = \prod_{k=1}^{L/2-n_\mp} f_1\left(\lambda_j^{(\pm)}-\lambda_k^{(\mp)}\right)\,
    f_1\left(\lambda_j^{(\pm)}+\lambda_k^{(\mp)}\right)\,,\\
    &\qquad\qquad j=1,\dots,\frac12 L-n_\pm\,
  \end{aligned}
\end{equation}
where we have defined
\begin{equation}
  \label{fdef}
  f_s(x) = \frac{x+is}{x-is}\,.
\end{equation}
The eigenvalues of the conserved $U(1)$ charges from the Cartan
subalgebra are determined by the numbers of Bethe roots.  The energy
of a state parameterized by a solution of (\ref{bae24}) is
\begin{equation}
  \label{ene-o24}
  E=L-1 - \sum_{j=1}^{L-n_1} a_{1/2}\left(\lambda_j^{(1)}\right)\,.
\end{equation}
with $a_s(x)=i\partial_x \ln f_s(x) = 2s/(x^2+s^2)$.
In the thermodynamic limit, $L\to\infty$, the root configurations
corresponding to the ground state and many low energy excitations are
found to consist of reals with finite $n_1$, $n_\pm$.  The ground
state of even length chains is realized in the sector with
$n_1=n_\pm=1$.  Solving the Bethe equations numerically and
extrapolating the finite size energies assuming a rational dependence
of the effective scaling dimension on $1/\log L$ we find
\begin{equation}
  X_{\mathrm{eff},0}^{(2|4)} (L) \simeq \frac18 - \frac7{16}\,\frac{1}{\log L}\, 
\end{equation}
corresponding to an effective central charge $c_\mathrm{eff}=-3$, as
expected from Eq.~(\ref{ceff-z}).

\begin{figure}[ht]
  \includegraphics[width=0.6\textwidth]{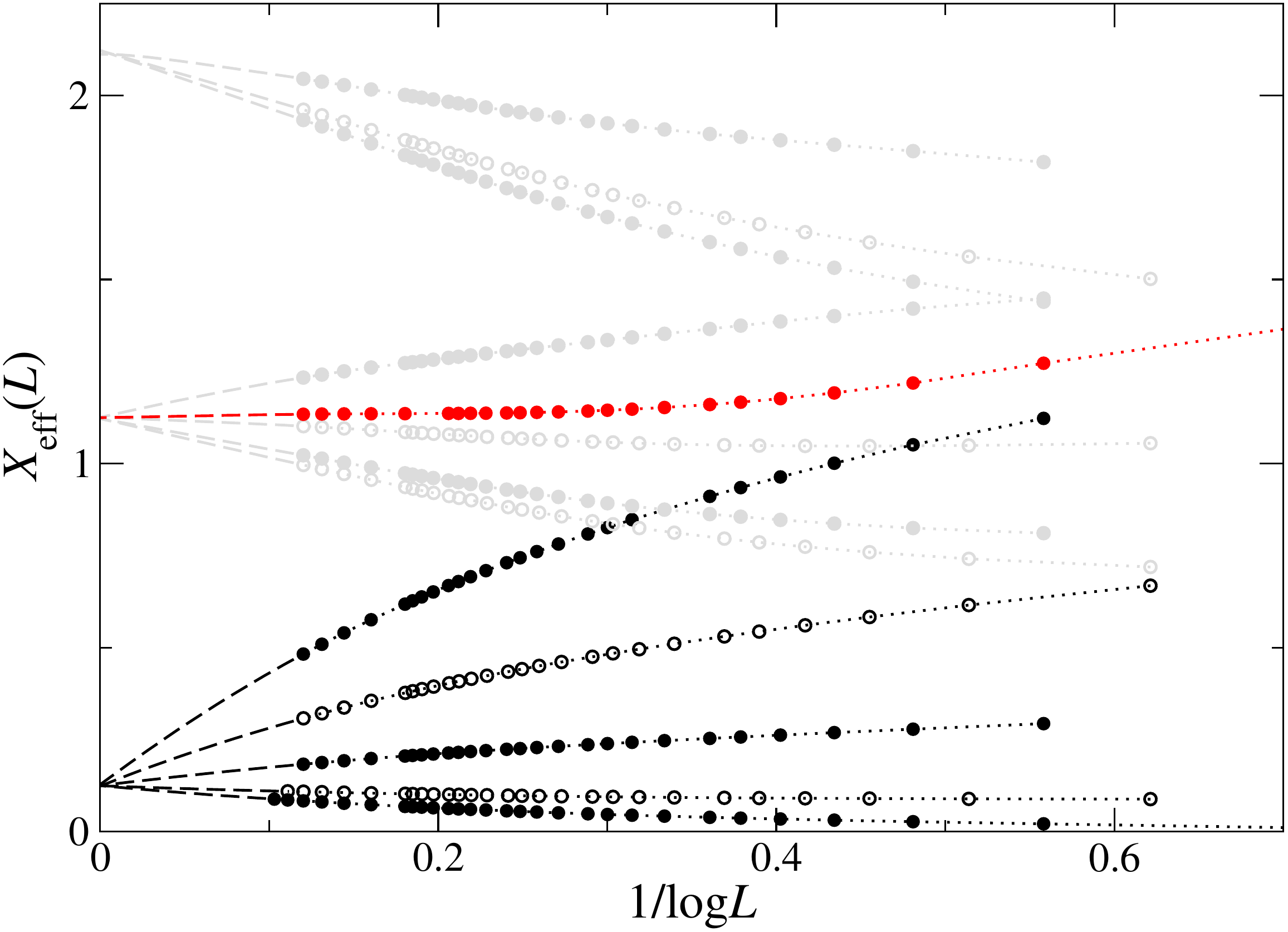}
  \caption{\label{fig:o24X} Corrections to the effective scaling dimensions $X_\mathrm{eff}=\frac18+n$ with $n=0$ (in black) and $1,2$ (in grey) for some of the low-lying states of the $OSp(2|4)$ chain. Data for even (odd) length are presented by filled (open) symbols, data shown in red correspond to a descendent state. Dashed lines are extrapolations to $L\to\infty$.}
\end{figure}

The lowest excitations appear in the sectors $n_1=1$,
$n_\pm=1\pm k/2$ with $|k|=1,2,3,\dots\sim L\mod 2$.  They, too, are
parameterized by real solutions to the Bethe equations
(\ref{bae24}). The leading finite size scaling of these states
coincides with that of the ground state, see Figure~\ref{fig:o24X}.
The subleading logarithmic corrections to scaling, however, vanish
with amplitudes depending on $|k|$ as
\begin{equation}
  \alpha^{(2|4)}(k) = \frac14 k^2-\frac7{16}\,.
\end{equation}
Note that these can be related to the corresponding amplitudes
observed in spectrum of the periodic $OSp(2|4)$ model,
$\beta^{(2|4)}(k) = (k^2-1)/8$ \cite{FrMa18}, by
\begin{equation}
  \alpha^{(2|4)}(k) = 2\beta^{(2|4)}(k) -\frac3{16}\,.
\end{equation}

Similar groups of excitations corresponding to primaries with scaling
dimension $X=1$ ($2$) appear in the sectors with $n_1=2$ and
$n_\pm = (3\pm k)/2$ for $|k|=0,1,2,\dots\sim (L+1)\mod 2$
($n_\pm=(2\pm k/2)$ for $|k|=0,1,2,\dots \sim L \mod 2$).

In addition we have identified the root configuration for an
excitation of the even length superspin chain in the sector
$n_1=n_\pm=1$: apart from the real roots it contains a pair of complex
conjugate roots $\lambda_{c\pm}^{(1)} \simeq \lambda_0 \pm i/2$ with
$\lambda_0\in \mathbb{R}^+$ on the first level and imaginary roots
$\lambda_c^{(+)}=-\lambda_c^{(-)}\simeq i/2$ (or $- i/2$) on the
second and third level.  Extrapolation of the finite size data gives
scaling dimensions $X=1$, see Figure~\ref{fig:o24X}, indicating that
this is a descendent of the ground state.

%
\subsection{$z=-1$: the $OSp(1|2)$ and $OSP(3|4)$ superspin chains}
%
\paragraph{$OSp(1|2)$.} Solutions of the Bethe equations
\begin{equation}
  \label{bae12}
  \begin{aligned}
    \left[f_{1/2}\left(\lambda_j\right)\right]^{2L} = \prod_{k\neq j}^{L-2n}
      f_1\left(\lambda_j-\lambda_k\right) 
      f_1\left(\lambda_j+\lambda_k\right) 
      f_{-1/2}\left(\lambda_j-\lambda_k\right) 
      f_{-1/2}\left(\lambda_j+\lambda_k\right) \,,
      \quad j=1,\dots,L-2n\,,
  \end{aligned}
\end{equation}
parameterize highest weight states for $(4n+1)$-dimensional
$OSp(1|2)$-multiplets with superspin $J=n$ where $2n$ is a
non-negative integer. The energy of this state is
\begin{equation}
  \label{ene-o12}
  E = (L-1) - \sum_{j=1}^{L-n} a_{1/2}\left(\lambda_j\right)\,.
\end{equation}

The lowest levels in each sector with given superspin $J>0$ are given
in terms of positive roots of the Bethe equations (\ref{bae12}). Among
these is the ground state of the $OSp(1|2)$ chain with both even and
odd length in the $J=1/2$ triplet sector.  From the extrapolation of
the finite size energies of this state we reproduce the known central
charge $c_{\mathrm{eff}}=-2$ for this model.  Up to subleading
corrections to scaling this state is degenerate with the $OSp(1|2)$
$J=0$ singlet with a root configuration consisting of $L-2$ positive
rapidities and a two-string of complex conjugate ones,
$\lambda_{c\pm}\simeq\lambda_0\pm i/2$ with
$\lambda_0\in \mathbb{R}^+$.  Complemented with results from the
finite size analysis of the ground states in the sectors $J>1/2$, we
find the conformal weights corresponding to the lowest states with
superspin $J$ to be
\begin{equation}
  X^{(1|2)}_J(L) \simeq J(2J-1)
  + \frac{\alpha^{(1|2)}(J)}{\log L}\,, \quad
  J=0,\frac12,1,\frac32,\dots\,.. 
\end{equation}  

We have identified the lowest excitation in the $J=1/2$ and $1$
sectors: the triplet excitation has a root configuration similar to
the $J=0$ ground state described above and corresponds to an operator
with conformal weight $X=1$. The excitation on top of the lowest $J=1$
state is given in terms of real roots with a particle-hole pair at the
Fermi point giving $X=2$.

The subleading corrections to scaling of some of these states have
been studied in Ref.~\cite{GrJS19}.  For the lowest states with
superspin $J>0$ they are
\begin{equation}
  \label{o12-alphaGJS}
  \alpha^{(1|2)}(J) = -\frac23 J(J+1) + \frac5{24}\,,\quad
  J=\frac12,1,\frac32,\dots\,.
\end{equation}

\paragraph{$OSp(3|4)$.} According to (\ref{specinclusion}) these
energies do appear in the spectrum of the $OSp(3|4)$ model.  The Bethe
equations for the latter (in grading $bffbffb$) are
\begin{equation}
  \label{bae34}
  \begin{aligned}
    &\left[f_{1/2}\left({\lambda_j^{(1)}}\right)\right]^{2L}
    = \prod_{k =1}^{N_2}
    f_{1/2}\left({\lambda_j^{(1)}-\lambda_k^{(2)}}\right)\,
    f_{1/2}\left({\lambda_j^{(1)}+\lambda_k^{(2)}}\right)\,,
    \quad j=1\ldots N_1\,,\\
    &\prod_{k=1}^{N_1}
    f_{1/2}\left(\lambda_j^{(2)}-\lambda_k^{(1)}\right)\,
    f_{1/2}\left(\lambda_j^{(2)}+\lambda_k^{(1)}\right)\,
    \prod_{k=1}^{N_3}
    f_{1/2}\left({\lambda_j^{(2)}-\lambda_k^{(3)}}\right)\,
    f_{1/2}\left({\lambda_j^{(2)}+\lambda_k^{(3)}}\right)\,\\
    &\qquad= \prod_{k\neq j}^{N_2}
    f_{1}\left({\lambda_j^{(2)}-\lambda_k^{(2)}}\right)\,
    f_{1}\left({\lambda_j^{(2)}+\lambda_k^{(2)}}\right)\,,
    \quad j=1\ldots N_2\,,\\
    &\prod_{k=1}^{N_2}
    f_{1/2}\left(\lambda_j^{(3)}-\lambda_k^{(2)}\right)\,
    f_{1/2}\left(\lambda_j^{(3)}+\lambda_k^{(2)}\right)\,\\
    &\qquad= \prod_{k\neq j}^{N_3}
    f_{1}\left({\lambda_j^{(3)}-\lambda_k^{(3)}}\right)\,
    f_{1}\left({\lambda_j^{(3)}+\lambda_k^{(3)}}\right)\,
    f_{-1/2}\left({\lambda_j^{(3)}-\lambda_k^{(3)}}\right)\,
    f_{-1/2}\left({\lambda_j^{(3)}+\lambda_k^{(3)}}\right)\,,
    \quad j=1\dots N_3\,.
  \end{aligned}
\end{equation}
The energy of a state corresponding to a root configuration of (\ref{bae34}) is
\begin{equation}
  \label{ene-o34}
  E = -(L-1) + \sum_{j=1}^{L-n} a_{1/2}\left(\lambda_j\right)\,.
\end{equation}

The root densities for the ground state and low energy excitations of
the $OSp(3|4)$ superspin chain are $N_i/L\to1$ in the thermodynamic
limit. As in Ref.~\cite{FrMa18} we label the charge sectors of this
model by quantum numbers
$(n_1,n_2,n_3) = (N_1-N_2+1, N_2-N_3+1, L-N_1-2)$.  For the low energy
states most Bethe roots are arranged in complex conjugate pairs as
\begin{equation}
  \label{o34-strings}
  \lambda^{(1)}_\pm \simeq \lambda^{(1)}\pm \frac{5i}{4}\,,\quad
  \lambda^{(2)}_\pm \simeq \lambda^{(2)}\pm \frac{3i}{4}\,,\quad
  \lambda^{(3)}_\pm \simeq \lambda^{(3)}\pm \frac{i}{4}\,,\quad
  \lambda^{a}\in \mathbb{R}^+\,.
\end{equation}

\begin{figure}[t]
  \includegraphics[width=0.6\textwidth]{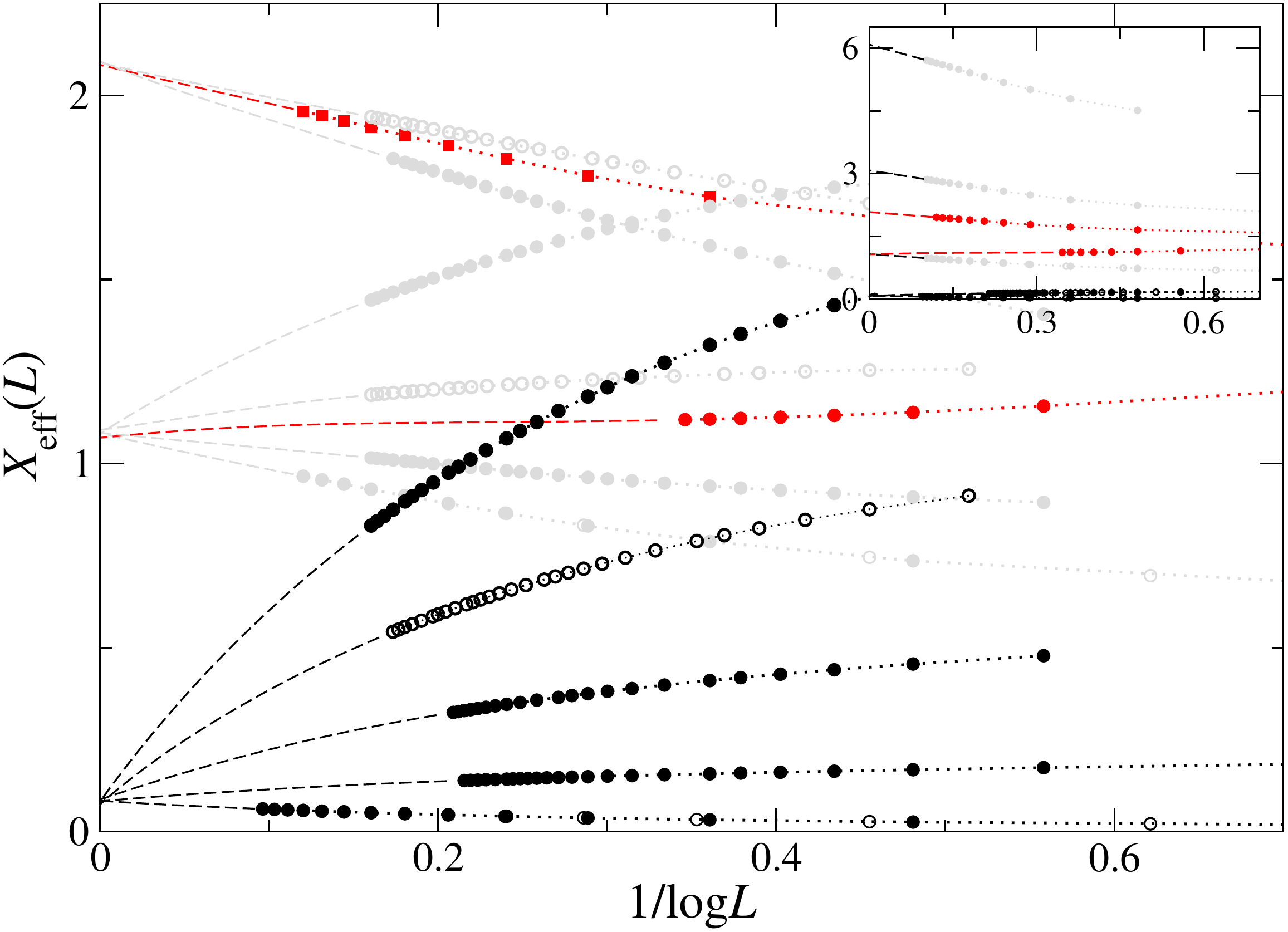}
  \caption{\label{fig:o34X} Corrections to the effective scaling dimensions $X_\mathrm{eff}=\frac{1}{12}+n$ for the lowest states of the $OSp(3|4)$ superspin chain. The inset shows the lowest levels of the $OSp(1|2)$ model. See Fig.~\ref{fig:o24X} for the meaning of symbols and colors.}
\end{figure}

The states with lowest energies states are parameterized by
configurations with $N_a=(L-k-2)/2$ of these strings, i.e.\ found in
the sectors $(n_1,n_2,n_3)=(1,1,k)$ with $k=0,1,2,\dots\sim L\mod 2$.
Among them the energies of the $k=0,1$ levels coincide with those of
the $J=1/2$ ground state and the lowest $J=0$ state of the $OSp(1|2)$
chain.  In the thermodynamic limit all of these states are degenerate
giving conformal weights $\lim_{L\to\infty}X^{(3|4)}_{(1,1,k)}(L)=0$,
see Figure~\ref{fig:o34X}.
From our numerical finite size data we find that this degeneracy is
lifted for finite $L$ by subleading corrections to scaling depending
on $k$ as
\begin{equation}
  \label{o34-alpha}
  X^{(3|4)}_{(1,1,k)}(L) \simeq \frac{\alpha^{(3|4)}(k)}{\log L}\,, \quad
  \alpha^{(3|4)}(k) = \frac13 k(k+1) - \frac7{24}\,, \quad
  k=0,1,2,\dots\,.
\end{equation}
Note that $\alpha^{(3|4)}(k=0)=\alpha^{(1|2)}(J=\frac12)$.  Comparing
these amplitudes to those for the model with periodic boundary
conditions \cite{FrMa18} we find
\begin{equation}
  \alpha^{(3|4)}(k) = 2\beta^{(3|4)}(k) -\frac18\,.
\end{equation}

A group of energies extrapolating to $X^{(3|4)}=1$ is found in the
sectors $(n_1,n_2,n_3)=(2,1,k)$ with $k=0,1,2,\dots\sim L+1\mod 2$.
Here one of the $N_1=L-2-k$ roots on the first level of (\ref{bae34})
is real.  The energy of the $k=0$ level coincides with the lowest
$J=1$ level of the $OSp(1|2)$ chain.  The next group of excitations
with $X^{(3|4)}=2$ is observed in the sectors $(n_1,n_2,n_3)=(2,2,k)$,
$k=0,1,2,\dots\sim L\mod 2$. A summary of the finite size spectrum of
the $OSp(1|2)$ and $OSp(3|4)$ is shown in Figure~\ref{fig:o34X}.

%
\subsection{$z=0$: the $OSp(2|2)$ superspin chain}
\label{sec:CFT-o22}
%
The Bethe equations for the $OSp(2|2)$ model (grading $fbbf$) are
\begin{equation}
  \label{bae22}
  \begin{aligned}
    &\left[f_{1/2}\left({\lambda_j^{(+)}}\right)\right]^{2L}
    = \prod_{k =1}^{N_-}
    f_{1}\left({\lambda_j^{(+)}-\lambda_k^{(-)}}\right)\,
    f_{1}\left({\lambda_j^{(+)}+\lambda_k^{(-)}}\right)\,,
    \quad j=1\ldots N_+\,,\\
    &\left[f_{1/2}\left({\lambda_j^{(-)}}\right)\right]^{2L}
    = \prod_{k =1}^{N_+}
    f_{1}\left({\lambda_j^{(-)}-\lambda_k^{(+)}}\right)\,
    f_{1}\left({\lambda_j^{(-)}+\lambda_k^{(+)}}\right)\,,
    \quad j=1\ldots N_-\,.
  \end{aligned}
\end{equation}
Each solution to these equations parameterizes an eigenstate of the
superspin chain with energy
\begin{equation}
  \label{ene-o22}
  E = (L-1) - \sum_{j=1}^{N_+} a_{1/2}\left(\lambda_j^{(+)}\right)
  - \sum_{j=1}^{N_-} a_{1/2}\left(\lambda_j^{(-)}\right)\,.
\end{equation}
The root configurations for the ground state and low energy
excitations of the model consist of real roots $\lambda_j^{(\pm)}>0$
with densities $N_\pm/L\to1/2$ in the thermodynamic limit. Using
quantum numbers $(n_1,n_2) = (L-N_+-N_-,N_+-N_-)$ for the $U(1)$
charges we find that the ground state of the model is realized in the
$(n_1,n_2)=(1,0)$ sector of the superspin chain with odd length.  The
effective central charge of the model is known to be
$c_{\mathrm{eff}}=-1$.  In the thermodynamic limit this state
degenerates with the lowest levels in the sectors $(n_1,n_2)=(1,k)$
for $k=1,2,3,\dots \sim L+1\mod 2$, all of them giving a conformal
weight $X_{(1,k)}^{(1|2)}=0$.  For finite $L$ the degeneracy is lifted
by logarithmic corrections to scaling
\begin{equation}
  X^{((2|2)}_{(1,k)}(L) \simeq \frac{\alpha^{(2|2)}(k)}{\log L}\,,
  \quad
  \alpha^{(2|2)}(k) = \frac12 k^2 - \frac5{16}\,.
\end{equation}
This expression can be related to that for the periodic $OSp(2|2)$
chain \cite{FrMa18} as follows
\begin{equation}
  \alpha^{(2|2)}(k) = 2\beta^{(2|2)}(k) -\frac1{16}\,.
\end{equation}

\begin{figure}[t]
  \includegraphics[width=0.6\textwidth]{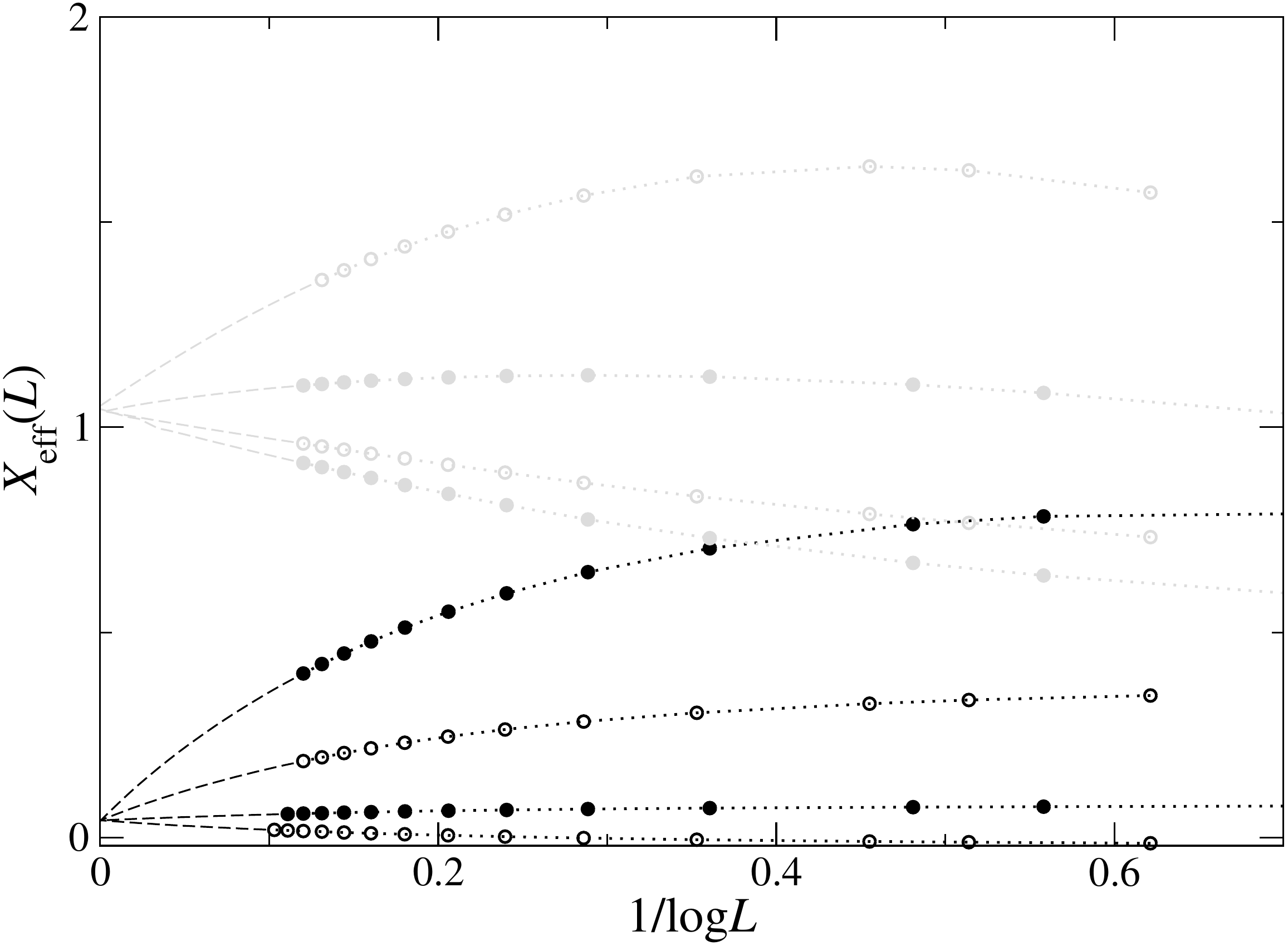}
  \caption{\label{fig:o22X} Corrections to the effective scaling dimensions $X_\mathrm{eff}=\frac{1}{24}+n$ for the lowest states of the $OSp(2|2)$ chain. See Fig.~\ref{fig:o24X} for the meaning of symbols and colors.}
\end{figure}

A similar tower of excitations giving conformal weight $X^{(2|2)}=1$
up to logarithmic corrections exists in the sectors $(n_1,n_2)=(2,k)$
for $k=0,1,2,\dots\sim L\mod 2$. The finite size scaling behaviour of
the states we have analyzed is presented in Figure~\ref{fig:o22X}.
%
\subsection{$z = 1$: the $OSp(3|2)$ superspin chain}
\label{sec:CFT-o32}
%
To study the finite size spectrum of the $OSp(3|2)$ superspin chain we
make use of the Bethe equations in two different gradings:
the Bethe equations for the model with open boundaries in the grading
$fbbbf$ read
\begin{equation}
  \label{bae32_fbbbf}
  \begin{aligned}
    &\left[f_{1/2}\left(\lambda_{j}^{(1)}\right)\right]^{{2L}}=
    \prod_{k=1}^{{L-n_1-n_2}}
    f_{1/2}\left({\lambda_{j}^{(1)}-\lambda_{k}^{(2)}}\right)\,
    f_{1/2}\left(\lambda_{j}^{(1)}+\lambda_{k}^{(2)}\right)\,,\quad j=1,\cdots,{L-n_1} , 
    \\
    &\prod_{k=1}^{{L-n_1}}
    f_{1/2}\left(\lambda_{j}^{(2)}-\lambda_{k}^{(1)}\right)\,
    f_{1/2}\left(\lambda_{j}^{(2)}+\lambda_{k}^{(1)}\right)=
    \\ &\qquad= 
    \prod_{k \ne j }^{{L-n_1-n_2}}
    f_{1/2}\left(\lambda_{j}^{(2)}-\lambda_{k}^{(2)}\right)\,
    f_{1/2}\left(\lambda_{j}^{(2)}+\lambda_{k}^{(2)}\right)\,, 
    \quad j=1,\cdots,{L-n_1-n_2} \,.
  \end{aligned}
\end{equation}
The corresponding energy is given in terms of the Bethe roots from the
first level as
\begin{equation}
  E_{fbbbf} = (L-1)
  - \sum_{j=1}^{L-n_1} a_{1/2}\left(\lambda_j^{(1)}\right)\,.
\end{equation}

Choosing the grading $bfbfb$ the spectrum of the open superspin chain
is parameterized by solutions to the Bethe equations:
\begin{equation}
  \label{bae32_bfbfb}
  \begin{aligned}
      &\left[f_{1/2}\left(\lambda_{j}^{(1)}\right)\right]^{{2L}}=
    \prod_{k=1}^{{L-n_1-n_2}}
    f_{1/2}\left({\lambda_{j}^{(1)}-\lambda_{k}^{(2)}}\right)\,
    f_{1/2}\left(\lambda_{j}^{(1)}+\lambda_{k}^{(2)}\right)\,,
    \quad j=1,\cdots,{L-n_2-1}, 
    \\
    &\prod_{k=1}^{{L-n_2-1}}
    f_{1/2}\left(\lambda_{j}^{(2)}-\lambda_{k}^{(1)}\right)\,
    f_{1/2}\left(\lambda_{j}^{(2)}+\lambda_{k}^{(1)}\right) =
    \\ &\qquad = 
    \prod_{k \ne j }^{{L-n_1-n_2}}
    f_{-1/2}\left(\lambda_{j}^{(2)}-\lambda_{k}^{(2)}\right) \,
    f_{-1/2}\left(\lambda_{j}^{(2)}+\lambda_{k}^{(2)}\right) \
    f_1\left(\lambda_{j}^{(2)}-\lambda_{k}^{(2)}\right))\,
    f_1\left((\lambda_{j}^{(2)}+\lambda_{k}^{(2)}\right)\,,\\ 
    & \qquad\qquad j=1,\cdots,{L-n_1-n_2} \,,
  \end{aligned}
\end{equation}
and the corresponding energy eigenvalue is
\begin{equation}
  E_{bfbfb} = -L+1 +
  \sum_{j=1}^{L-n_2-1} a_{1/2}\left(\lambda_j^{(1)}\right)\,.
\end{equation}

Solutions to the Bethe equations (\ref{bae32_fbbbf}) and
(\ref{bae32_bfbfb}) parameterize highest weight states of $OSp(3|2)$
in the irreducible representations $(p;q)$ appearing in the tensor
product $(0;\frac12)^{\otimes L}$ of local spins
\cite{Jeugt84,FrMa15}. In terms of the number of Bethe roots the
quantum numbers $p$ and $q$ are given as
\begin{equation}
    p=n_1-1\,,\quad q=(n_2+1)/2\,.
\end{equation}

Exact diagonalization of the $OSp(3|2)$ Hamiltonian shows that the
ground state is a $(p;q)=(0;0)$ singlet ($(0;\frac12)$ quintet) for
$L$ even (odd).  Its energy is
$E_0=L e_\infty + f_\infty \equiv-3(L-1)$ without any finite size
corrections -- similar to the model with periodic boundary conditions
-- giving the effective central charge $c_{\mathrm{eff}}=0$.. The
$fbbbf$ Bethe root configuration for $L$ odd contains $(L-1)/2$ pairs
of complex conjugate rapidities
$\lambda^{(a)}_{j\pm}\simeq\lambda^{(a)}_j\pm i/4$ with positive
$\lambda^{(a)}_j$ on each level $a=1,2$.  The root configuration for
even $L$ contains degenerate roots.

\begin{figure}[t]
  \includegraphics[width=0.6\textwidth]{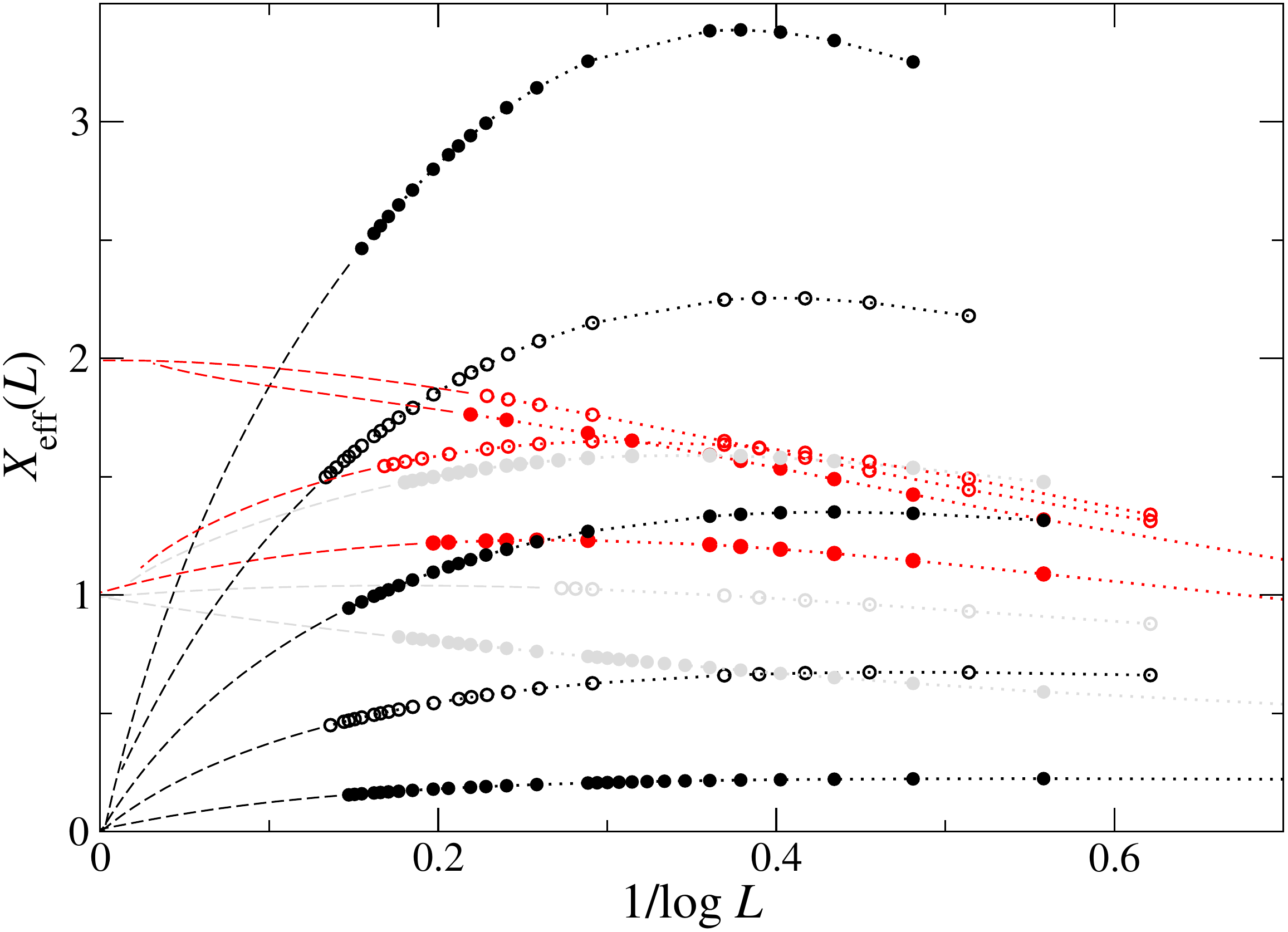}
  \caption{\label{fig:o32X} Corrections to the effective scaling dimensions $X_\mathrm{eff}=n$ for the lowest excitations of the $OSp(3|2)$ chain in the sectors $(0;q)$ and $(1;q)$ (the ground state in the sectors $(0;0)$, $(0;\frac12)$ with $X_\mathrm{eff}\equiv0$ independent of $L$ is not shown). See Fig.~\ref{fig:o24X} for the meaning of symbols and colors.}
\end{figure}

As for the models considered above the finite size spectrum of the
$OSp(3|2)$ superspin chain can be grouped into sets extrapolating to
the same integer conformal weight in the thermodynamic limit.
Specifically, the lowest states in the sectors $(0;q)$ with
$2q=0,1,2,\dots$ (or $(n_1,n_2)=(1,k)$ with integer $k$) become
degenerate with the ground state, see Figure \ref{fig:o32X}.
At large but finite $L$ this degeneracy is lifted by logarithmic corrections to
scaling
\begin{equation}
  X^{(3|2)}_{(0;q)}(L) \simeq \frac{\alpha^{(3|2)}(q)}{\log L}\,,\quad
  \alpha^{(3|2)}(q) = 2q(2q-1)\,.
\end{equation}
We note that the amplitudes $\alpha^{(3|2)}$ are twice of those found for the periodic $OSp(3|2)$ chain \cite{FrMa15}, i.e.\ $\alpha^{(3|2)}(q) = 2\beta^{(3|2)}(q)$.

Finite size data for the lowest states in the sectors $(p;q)=(1;q)$,
$q=\frac12,1,\frac32$, and some descendents states are also shown in
Figure \ref{fig:o32X}.

\section{Discussion}

In this paper we have investigated the finite-size properties 
of the spectrum of the
$OSp(n|2m)$ superchain chain with free boundary conditions. We perform this analysis
by solving numerically the corresponding Bethe equations for large systems sizes. This
study made it possible to 
identify the corresponding operator content and to extract the amplitudes
associated to the subleading corrections to the asymptotic behaviour.  

For $z=n-2m <2$ we find that the surface exponents are built out of a set of integer numbers. Similar as in the case of periodic boundary conditions this was to be expected based on the perturbative RG analysis of the model.  The surface exponents turn out to be exactly the same as the bulk exponents which is a peculiarity of the underlying universality class.
The spectra contain an abundance of states with null conformal dimension whose degeneracy is lifted by subleading logarithmic corrections. We find that the amplitudes of such corrections are different for periodic and free boundary conditions. From our numerical analysis we conjecture a simple relation among these amplitudes to be
\begin{equation}
\alpha(k)= 2\beta(k) +\frac{z-1}{16}\,,\qquad z < 2\,,
\end{equation}
where $\alpha(k)$ and $\beta(k)$ correspond to the amplitudes associated to free and periodic boundaries, respectively. 

We now can use this result to infer about the asymptotic behaviour of the 
surface watermelon correlators associated to the respective loop model. The above
relation tells us that we can use the same reference state $k_0$ associated 
to the smallest
non-negative amplitudes for both free and periodic boundaries. Proceeding in analogy
as has already been explained for periodic boundary \cite{FrMa18} we conjecture that the
surface correlators should behave as
\begin{equation}
G^{(s)}_{k}(\rho) \sim 1/\ln(\rho)^{2\gamma(k)}\,,\qquad\gamma(k)= \frac{k(k+z-2)}{2-z}\,,
\end{equation}
where $\rho$ denotes the distance among two points close to the boundary.

We conclude by recalling some existing results on the finite-size properties in the regime $z \geq 2$ for free boundary conditions. For the $O(2)$ (spin-$1/2$ Heisenberg) model of even length it has been argued that all conformal dimensions are given by the identity conformal tower \cite{ABGR88,EgAf92}. The logarithmic corrections to scaling for this model have been computed in \cite{AfQi99}.  In particular the gap between the ground state and the lowest (triplet) excitation is given by
\begin{equation}
E_1^{O(2)}(L)-E_0^{O(2)}(L) \simeq \frac{\pi v_F}{L}\left(1-1/\ln(L)\right)\,,
\end{equation}
corresponding to conformal weight $X=1$ where the amplitude of the logarithmic correction to scaling is determined by the quadratic Casimir of the underlying algebra.
The spectrum of the $O(4)$ chain can be composed from the eigenenergies of two decoupled Heisenberg chains \cite{MaRa97a}. This can be used to infer the corresponding behaviour, in particular (note that the Fermi velocities of the $O(2)$ and the $O(4)$ model coincide, see Table~\ref{tab1})
\begin{equation}
E_1^{O(4)}(L)-E_0^{O(4)}(L) \simeq \frac{\pi v_F}{L}\left(1-1/\ln(L)\right)\,.
\end{equation}
%

Based on our previous work on the $OSp(5|2)$ superspin chain with periodic boundary conditions \cite{FrMa18} we expect towers of levels with the same conformal weight to be present in the regime $z=n-2m \geq 2$ but $m>0$. 
We hope to investigate how these degeneracies are lifted by subleading corrections to scaling in a forthcoming paper.

\begin{acknowledgments}
Partial support for the work of HF is provided by the Deutsche Forschungsgemeinschaft under grant No.\ Fr~737/9-2 within the research unit   \emph{Correlations in Integrable Quantum Many-Body Systems} (FOR2316).
The work of MJM was supported in part by the Brazilian 
Research Council CNPq
through the grants 304758/2017-7.
\end{acknowledgments}

\appendix
\section{The surface energy}
\label{app:thermo}
To employ Eq.~(\ref{EFRE}) for the analysis of the finite size spectrum of the open $OSp(n|2m)$ superspin chains with free boundary conditions the corresponding surface energy is needed as an input.  As a consequence of the spectral inclusion (\ref{specinclusion}) it is sufficient to consider the cases based on the superalgebras $OSp(1|2m)$, $OSp(2|2m)$ and $OSp(3|2)$ for the regime $z=n-2m<2$ considered in the main text. Here the last of these is special due to the singular Bethe root configuration for the ground state, see Section~\ref{sec:CFT-o32}.  Based on our numerical results for  $OSp(3|2)$ superspin chains of finite length, however, we conclude that the surface energy for this model is $f_\infty=3$.
The $\mathcal{O}(L^0)$ contribution to the ground state energy for the other two series of models can be computed using the root density method \cite{YaYa69,HaQB87}.  We start our discussion below by presenting the main steps of this approach for the $OSp(2|2)$ superspin chain which, as discussed in Section~\ref{sec:CFT-o22}, is solved by a nested Bethe ansatz with real roots for the low energy states. Using similar arguments we then apply this method to derive the surface energies for the spin chains based on $OSp(2|2m)$, $OSp(1|2m)$.

For sake of completeness we also compute the surface energies for the spin chains in the regime $z>2$ (the case $z=2$ can be represented by the isotropic spin-$1/2$ Heisenberg magnet (\ref{XXX}) as discussed in the main text).  For the models based on the ordinary Lie algebras $O(2n)$ the root density approach can be applied as before.  In the case of the $O(2n+1)$ spin chains it has to be modified slightly due to the presence of complex roots in the ground state configuration.  For the $O(3)$ and $O(5)$ model we will show at the end of this Appendix, that this can be dealt with using the so-called string hypothesis.

The results obtained in this appendix are summarized in Eqs.~(\ref{finf}) and (\ref{finf2}).

\subsection{$OSp(2|2)$ in the grading $fbbf$}
\label{app:finf-o22}
We start by taking the logarithm 
of the Bethe ansatz equations (\ref{bae22})
associated to
the $OSp(2|2)$ model for configurations of real roots $\lambda_j^{(\pm)}$. As a result we find
\begin{equation}
\label{bae22log}
\begin{aligned}
& 2L \phi_{1/2}\left(\lambda_j^{(+)}\right)= 2\pi \mathrm{Q}_j^{(+)}+ \sum_{k=1}^{N_{-}} \left[\phi_1\left(\lambda_j^{(+)}-\lambda_k^{(-)}\right)
+\phi_1\left(\lambda_j^{(+)}+\lambda_k^{(-)}\right) \right],~~~j=1,\dots,N_+\,,\\
& 2L \phi_{1/2}\left(\lambda_j^{(-)}\right)= 2\pi \mathrm{Q}_j^{(-)}+ \sum_{k=1}^{N_{+}} \left[\phi_1\left(\lambda_j^{(-)}-\lambda_k^{(+)}\right)
+\phi_1\left(\lambda_j^{(-)}+\lambda_k^{(+)}\right) \right],~~~j=1,\dots,N_{-}\,,
\end{aligned}
\end{equation}
a1where $\phi_s(x)\equiv2 \arctan(x/s)$ and the numbers $\mathrm{Q}_j^{(\pm)}$ are positive integers characterizing the possible branches of the logarithm.

From the above Bethe equations the so-called counting functions \cite{YaYa69}
\begin{equation}
\label{COUNT}
\begin{aligned}
& z_L^{(+)}(\lambda)=\frac{\phi_{1/2}(\lambda)}{\pi}  
-\frac{1}{2\pi L} \sum_{k=1}^{N_{-}} \left[\phi_1\left(\lambda-\lambda_k^{(-)}\right)
+\phi_1\left(\lambda+\lambda_k^{(-)}\right) \right]\,,\\
& z_L^{(-)}(\lambda)=\frac{\phi_{1/2}(\lambda)}{\pi}   
-\frac{1}{2\pi L} \sum_{k=1}^{N_{+}} \left[\phi_1\left(\lambda-\lambda_k^{(+)}\right)
+\phi_1\left(\lambda+\lambda_k^{(+)}\right) \right] \,,
\end{aligned}
\end{equation}
take values $z_L^{(\pm)}\left(\lambda_j^{(\pm)}\right)=\mathrm{Q}_j^{(\pm)}/L$ for $j=1,\dots,N_\pm$.  For the lowest states of the towers considered in Sect.~\ref{sec:CFT-o22} we have $N_+=N_-\lesssim L/2$ with uniformly spaced quantum numbers $\mathrm{Q}_j^{(\pm)}=j$. Hence densities of the Bethe roots $\lambda_j^{(\pm)}$ in these states can be derived from the counting functions as
\begin{equation}
\label{rhoFI}
  \rho_{L}^{(\pm)}(\lambda)= \frac{d}{d \lambda}  z_L^{(\pm)}(\lambda)
    = \frac{a_{1/2}(\lambda)}{\pi} + \frac1{2\pi L} a_1(\lambda) 
     -\frac{1}{2\pi L} \sum_{k=-N_{\mp}}^{N_{\mp}} a_1\left(\lambda-\lambda_k^{(\mp)}\right) \,
\end{equation}
where we have symmetrized the sums by extending the sets of roots to $\{\lambda_j^{(\pm)}\} \cup \{0\} \cup \{\lambda_{-j}^{(\pm)}\equiv -\lambda_j^{(\pm)}\}$.
Note that these relations are similar to those obtained for periodic boundary conditions except for the presence of the additional boundary terms $a_1(\lambda)/(2\pi L)$. We anticipate that these terms are responsible to provide the surface contribution to the ground state energy.

For $L\gg 1$ the extended set of Bethe roots for the ground state tends to a continuous distribution on the entire real axis with densities $\rho_0^{(\pm)}(\lambda)$ and the sums in (\ref{rhoFI}) can be replaced by integrals
\begin{equation}
\label{rhoINF}
\begin{aligned}
& \rho_0^{(+)}(\lambda)\simeq\frac{a_{1/2}(\lambda)}{\pi} + \frac1{2\pi L} a_1(\lambda)
-\frac{1}{2\pi} \int_{-\infty}^{+\infty}  \mathrm{d} \mu\, a_1\left(\lambda-\mu\right) 
\rho_0^{(-)}(\mu)\,, \\
& \rho_0^{(-)}(\lambda)\simeq\frac{a_{1/2}(\lambda)}{\pi} + \frac1{2\pi L} a_1(\lambda)
-\frac{1}{2\pi} \int_{-\infty}^{+\infty}  \mathrm{d} \mu\, a_1\left(\lambda-\mu\right) 
\rho_0^{(+)}(\mu)\,.
\end{aligned}
\end{equation}
These integral equations can be solved order by order in powers of $L^{-1}$ by elementary Fourier techniques resulting in $\rho^{(\pm)}(\omega) = \sigma_0(\omega) + \tau_0(\omega)/L$ with
\begin{equation}
\label{FOUR2}
\sigma_0(\omega) = \frac1{\cosh(\omega/2)}\,, \quad
\tau_0(\omega) = \frac{\exp(-|\omega|)}{1+\exp(-|\omega|)}\,.
\end{equation}
Similarly, we rewrite the ground state energy (\ref{ene-o22}) as
\begin{equation}
    E_0/L \simeq
        1 - \frac12 \int_{-\infty}^\infty a_{1/2}(\lambda) \left(\rho_0^{(+)}(\lambda) + \rho_0^{(-)}(\lambda)\right)
        +\frac1L \left(-1 + a_{1/2}(0)\right)\,.
\end{equation}
Using (\ref{FOUR2}) we reproduce the known result $\epsilon_\infty=1-4\ln 2$ for the bulk energy density and obtain
\begin{equation}
    f_\infty = -1+a_{1/2}(0) - 2\int_{0}^{\infty} \mathrm{d}\lambda\, a_{1/2}(\lambda) \tau_0(\lambda)
    = 3-\int_{-\infty}^\infty \mathrm{d}\omega\, \mathrm{e}^{-|\omega|/2} \tau_0(\omega) = \pi-1\,
\end{equation}
for the surface energy of the $OSp(2|2)$ spin chain as shown in Table~\ref{tab1}.

\subsection{$OSp(2|2m)$ on the grading $f\dots fbbf \dots f$}
For this model the ground state and the low-lying excitations 
are described in terms real roots in the
$f\dots fbbf \dots f$ basis ordering. The Bethe equations
are given by
\begin{equation}
\label{betheOSP22m}
\begin{aligned}
\delta_{\ell,1} (2L) \phi_{1/2}\left(\lambda_{j}^{(\ell)}\right)=&2 \pi \mathrm{Q}_{j}^{(\ell)}
+\sum_{\stackrel{k=1}{k \neq j}}^{N_{\ell}} \left[
\phi_{1}\left(\lambda_{j}^{(\ell)} - \lambda_{k}^{(\ell)}\right) 
+\phi_{1}\left(\lambda_{j}^{(\ell)} + \lambda_{k}^{(\ell)}\right) \right] \\
&-\sum_{\alpha=\ell\pm 1} \sum_{k=1}^{N_{\alpha}} \left[
\phi_{1/2}\left(\lambda_{j}^{(\ell)} - \lambda_{k}^{(\alpha)}\right) 
+\phi_{1/2}\left(\lambda_{j}^{(\ell)} + \lambda_{k}^{(\alpha)}\right) \right]
,\quad \ell=1,\dots,m-2 \,, \\
\delta_{m,2} (2L) \phi_{1/2}\left(\lambda_{j}^{(m-1)}\right)=&2 \pi \mathrm{Q}_{j}^{(m-1)}
+\sum_{\stackrel{k=1}{k \neq j}}^{N_{m-1}} \left[
\phi_{1}\left(\lambda_{j}^{(m-1)} - \lambda_{k}^{(m-1)}\right)
+\phi_{1}\left(\lambda_{j}^{(m-1)} + \lambda_{k}^{(m-1)}\right) \right]  \\
&-\sum_{{\alpha=m -2,\pm}} \sum_{k=1}^{N_{\alpha}} \left[
\phi_{1/2}\left(\lambda_{j}^{(m-1)} - \lambda_{k}^{(\alpha)}\right)
+\phi_{1/2}\left(\lambda_{j}^{(m-1)} + \lambda_{k}^{(\alpha)}\right) \right]\,, \\
\delta_{m,1} (2L) \phi_{1/2}\left(\lambda_{j}^{(\ell)}\right) =& 2 \pi \mathrm{Q}_{j}^{(\pm)}
+\sum_{k=1}^{N_{\mp}} \left[
\phi_{1}\left(\lambda_{j}^{(\ell)} - \lambda_{k}^{(\mp)}\right) 
+\phi_{1}\left(\lambda_{j}^{(\ell)} + \lambda_{k}^{(\mp)}\right) \right] \\
&-\sum_{k=1}^{N_{m-1}} \left[
\phi_{1/2}\left(\lambda_{j}^{(\ell)} - \lambda_{k}^{(m-1)}\right) 
+\phi_{1/2}\left(\lambda_{j}^{(\ell)} + \lambda_{k}^{(m-1)}\right) \right] 
,\quad \ell=\pm \,, \\
\end{aligned}
\end{equation}
and the corresponding energy is
\begin{equation}
\label{ene-o22m}
    E=L-1-\sum_{j} a_{1/2}(\lambda^{(1)}_j)\,.
\end{equation}
The ground state is parameterized by $N_1=\dots=N_{m-1}= L$ and $N_\pm = L/2$ roots distributed on the positive real axis  in the thermodynamic limit.  Hence, by proceeding as for the $OSp(2|2)$ model above we obtain the $L^{-1}$ boundary contributions $\tau_0^{(\ell)}(\lambda)$, $\ell=1,\dots,m-1,\pm$ to the densities.  The energy (\ref{ene-o22m}) of the $OSp(2|2m)$ superspin chain is given in terms of the first level roots, $\ell=1$. The Fourier representation of their boundary density is
\begin{equation}
    \tau_0^{(1)}(\omega) = -2\mathrm{e}^{-(m-1)|\omega|/4}\,\frac{\sinh(m|\omega|/4)\cosh(\omega/4)}{\cosh(m\omega/2)}\,.
\end{equation}
From (\ref{ene-o22m}) the resulting surface energy is given in terms of $\tau_0^{(1)}$ as
\begin{equation}
\label{FREEosp22m}
    f_\infty = -1+\frac12 a_{1/2}(0) - \int_{0}^{\infty} \mathrm{d}\lambda\, a_{1/2}(\lambda) \tau_0^{(1)}(\lambda)
    = 1-\int_{0}^\infty \mathrm{d}\omega\, \mathrm{e}^{-|\omega|/2} \tau_0^{(1)}(\omega)
\end{equation}
After some manipulations with the help of the identity
\begin{equation}
\label{betaf}
\int_{0}^{+\infty}
\frac{\exp(-\mu x)}{\cosh(x)}  dx =\frac{1}{2}\Big(\psi(\mu/2+1/2)-\psi(\mu/2) \Big)\,,
\end{equation}
we can rewrite the surface free energy of the $OSp(2|2m)$ superspin chain in terms 
of the Euler $\psi$ function
as presented in the main text (\ref{finf}) with $z=2-2m$.

\subsection{$OSp(1|2m)$ on the grading $f\dots fbf \dots f$}
In the $f\dots fbf \dots f$ grading the ground state and the low-lying excitations of this model
are described in terms of positive rapidities satisfying the Bethe equations
\begin{equation}
\label{betheOSP12m}
\begin{aligned}
\delta_{\ell,1} (2L) \phi_{1/2}\left(\lambda_{j}^{(\ell)}\right)=&2 \pi \mathrm{Q}_{j}^{(\ell)}
+\sum_{\stackrel{k=1}{k \neq j}}^{N_{\ell}} \left[
\phi_{1}\left(\lambda_{j}^{(\ell)} - \lambda_{k}^{(\ell)}\right) 
+\phi_{1}\left(\lambda_{j}^{(\ell)} + \lambda_{k}^{(\ell)}\right) \right]  \\
&-\sum_{\alpha=\ell\pm 1} \sum_{k=1}^{N_{\alpha}} \left[
\phi_{1/2}\left(\lambda_{j}^{(\ell)} - \lambda_{k}^{(\alpha)}\right) 
+\phi_{1/2}\left(\lambda_{j}^{(\ell)} + \lambda_{k}^{(\alpha)}\right) \right]
\,,\quad \ell=1,\dots,m-1\,, \\
\delta_{m,1} (2L) \phi_{1/2}\left(\lambda_{j}^{(m)}\right) =& 2 \pi \mathrm{Q}_{j}^{(m)}
+\sum_{\stackrel{k=1}{k \neq j}}^{N_{m}} \left[
\phi_{1}\left(\lambda_{j}^{(m)} - \lambda_{k}^{(m)}\right) 
+\phi_{1}\left(\lambda_{j}^{(m)} + \lambda_{k}^{(m)}\right) 
\right]  \\
&- \sum_{\stackrel{k=1}{k \neq j}}^{N_{m}} \left[
\phi_{1/2}\left(\lambda_{j}^{(m)} - \lambda_{k}^{(m)}\right) 
+\phi_{1/2}\left(\lambda_{j}^{(m)} + \lambda_{k}^{(m)}\right) 
\right] \\
&-\sum_{k=1}^{N_{m-1}} \left[
\phi_{1/2}\left(\lambda_{j}^{(m)} - \lambda_{k}^{(m-1)}\right) 
+\phi_{1/2}\left(\lambda_{j}^{(m)} + \lambda_{k}^{(m-1)}\right) \right] 
\,. \\
\end{aligned}
\end{equation}
The energy associated to a solution is given again by (\ref{ene-o22m}).
In the thermodynamic limit we can introduce densities to describe the root configuration. Solving the corresponding integral equations as above the Fourier expression for the boundary contribution $\tau_0^{(1)}$ to the density of first level roots is found to be
\begin{equation}
    \tau_0^{(1)}(\omega) = -2\mathrm{e}^{-(2m-1)|\omega|/8}\,\frac{\sinh((2m+1)|\omega|/8)\cosh(\omega/4)}{\cosh((2m+1)\omega/4)}\,.
\end{equation}
The surface energy of the $OSp(1|2m)$ superspin chain can be computed from (\ref{FREEosp22m}) which,
using (\ref{betaf}), can be brought into the form  (\ref{finf}) with $z=1-2m$.

\subsection{$O(2n)$}
\label{app:thermoO2n}
As has been discussed in the main text the thermodynamical properties including the surface energy of the $O(2)$ model are known from the studies of the isotropic spin $s=1/2$ Heisenberg model (\ref{XXX}).
Similarly, the spectrum of the $O(4)$ spin chain can be derived by composing the eigenenergies of two decoupled Heisenberg spin chains \cite{MaRa97a}. From this identification we obtain that the surface free energy of the $O(4)$ spin chain is
\begin{equation}
f_{\infty}^{O(4)}= 2f_{\infty}^{O(2)}-1= \pi-2\ln(2)-1\,.
\end{equation}

From now on we shall concentrate our analysis for the models with $ n \geq 3$. The corresponding Bethe equations are given by
\begin{equation}
\label{betheO2n}
\begin{aligned}
\delta_{\ell,1} (2L) \phi_{1/2}\left(\lambda_{j}^{(\ell)}\right)=&2 \pi \mathrm{Q}_{j}^{(\ell)}
+\sum_{\stackrel{k=1}{k \neq j}}^{N_{l}} \left[
\phi_{1}\left(\lambda_{j}^{(\ell)} - \lambda_{k}^{(\ell)}\right) 
+\phi_{1}\left(\lambda_{j}^{(\ell)} + \lambda_{k}^{(\ell)}\right) \right]  \\
&-\sum_{\alpha=l\pm 1} \sum_{k=1}^{N_{\alpha}} \left[
\phi_{1/2}\left(\lambda_{j}^{(\ell)} - \lambda_{k}^{(\alpha)}\right) 
+\phi_{1/2}\left(\lambda_{j}^{(\ell)} + \lambda_{k}^{(\alpha)}\right) \right]
\,,\quad \ell=1,\dots,n-3 \,, \\
\delta_{n,3} (2L) \phi_{1/2}\left(\lambda_{j}^{(n-2)}\right)=&2 \pi \mathrm{Q}_{j}^{(n-2)}
+\sum_{\stackrel{k=1}{k \neq j}}^{N_{l}} \left[
\phi_{1}\left(\lambda_{j}^{(n-2)} - \lambda_{k}^{(n-2)}\right)
+\phi_{1}\left(\lambda_{j}^{(n-2)} + \lambda_{k}^{(n-2)}\right) \right] \\
&-\sum_{\alpha=n-3,\pm} \sum_{k=1}^{N_{\alpha}} \left[
\phi_{1/2}\left(\lambda_{j}^{(n-2)} - \lambda_{k}^{(\alpha)}\right)
+\phi_{1/2}\left(\lambda_{j}^{(n-2)} + \lambda_{k}^{(\alpha)}\right) \right]\,, \\
\delta_{n,2} (2L) \phi_{1/2}\left(\lambda_{j}^{(\ell)}\right) =& 2 \pi \mathrm{Q}_{j}^{(\pm)}
+\sum_{\stackrel{k=1}{k \neq j}}^{N_{\pm}} \left[
\phi_{1}\left(\lambda_{j}^{(\ell)} - \lambda_{k}^{(\ell)}\right) 
+\phi_{1}\left(\lambda_{j}^{(\ell)} + \lambda_{k}^{(\ell)}\right) \right] \\
&-\sum_{k=1}^{N_{n-2}} \left[
\phi_{1/2}\left(\lambda_{j}^{(\ell)} - \lambda_{k}^{(n-2)}\right) 
+\phi_{1/2}\left(\lambda_{j}^{(\ell)} + \lambda_{k}^{(n-2)}\right) \right] 
\,,\quad \ell=\pm\,. \\
\end{aligned}
\end{equation}
The energy of the corresponding eigenstate of the $O(2n)$ model is given again in terms of the first level roots by the expression (\ref{ene-o22m}).
The ground state and low lying excitations are parameterized by real roots $\lambda_j^{(\ell)}>0$ with total densities $n_\ell=N_\ell/L=1$ and $n_\pm=N_\pm/L=1/2$ in the thermodynamic limit.  Proceeding as for the models discussed in the previous sections we obtain the Fourier expression for the boundary contribution $\tau_0^{(1)}$ to the density of roots $\lambda_j^{(1)}$:
\begin{equation}
    \tau_0^{(1)}(\omega) = -\mathrm{e}^{-(n-2)|\omega|/4}\,
      \frac{\sinh(n|\omega|/4)-\cosh((n-2)\omega/4)}{\cosh((n-1)\omega/2)}\,.
\end{equation}
Using this expression in (\ref{FREEosp22m}) we find that the surface energy of the $O(2n)$ spin chain is given by (\ref{finf2}) with $z=2n$.

\subsection{$O(3)$}
The Bethe equations for the integrable $O(3)$ spin chain (or, equivalently, the spin $S = 1$ Takhtajan-Babujian model \cite{Takh82,Babu82}) read
\begin{equation}
    \label{baeO3}
      \left[f_{1/2}(\lambda_j)\right]^{2L}= \prod_{k\ne j}^{L-n} f_{1/2}(\lambda_j-\lambda_k) f_{1/2}(\lambda_j+\lambda_k)\,,
        \quad j=1,\dots,L-n\,.
\end{equation}
The corresponding energy eigenvalue is given as
\begin{equation}
    \label{ene-o3}
    E= L-1 -\sum_j^{L-n} a_{1/2}(\lambda_j)
    \,.
\end{equation}
The ground state of the model for even $L$ is parametrized by a solution of (\ref{baeO3}) in the sector $n=0$ containing $L/2$ two-strings $x_{j,\pm}\simeq \xi_j\pm i/4$, $\xi_j>0$.  Neglecting corrections to the imaginary parts the Bethe equations can be rewritten in terms of the coordinates of the string centers. Taking the logarithm we obtain
\begin{equation}
\begin{aligned}
    &2L \left( \phi_{3/4}(\xi_j) + \phi_{1/4}(\xi_j) \right) = 2\pi \mathrm{Q}_j - \phi_{1/2}(\xi_j)\\
    &\qquad 
    - \sum_{k=1}^{L/2} \left[ 2\left( \phi_{1/2}(\xi_j-\xi_k) + \phi_{1/2}(\xi_j+\xi_k) \right)
                             + \phi_{1}(\xi_j-\xi_k) + \phi_{1}(\xi_j+\xi_k) \right]
\end{aligned}
\end{equation}
with quantum numbers $\mathrm{Q}_j=1,2,\dots,L/2$.  Similarly, the energy (\ref{ene-o3}) becomes
\begin{equation}
    E_0 = L-1 - \sum_{j=1}^{L/2} \left( a_{1/4}(\xi_j) + a_{3/4}(\xi_j) \right) \,.
\end{equation}
Proceeding as in Appendix~\ref{app:finf-o22} we obtain an integral equation for the density of strings in the ground state for $L\gg1$
\begin{equation}
\begin{aligned}
    \rho_0(\xi) \simeq& \frac1\pi \left(a_{3/4}(\xi) + a_{1/4}(\xi)\right)
                + \frac1{2\pi L}\left(3a_{1/2}(\xi) +a_1(\xi)\right)\\
                &-\frac1{2\pi}\int_{-\infty}^\infty\mathrm{d}\xi'\,
                   \left[2a_{1/2}(\xi-\xi')+a_1(\xi-\xi')\right] \rho_0(\xi')\,.
\end{aligned}
\end{equation}
Solving this integral equation by Fourier methods we obtain
\begin{equation}
    \tau_0(\omega) = \frac{3+\exp(-|\omega|/2)}{4\cosh^2(\omega/4))}
\end{equation}
for the $\mathcal{O}(L^{-1})$ contribution to the density.  Hence we have
\begin{equation}
\begin{aligned}
    f_\infty &= -1+\frac12\left(a_{1/4}(0)+a_{3/4}(0)\right)
               - \int_0^{\infty} \mathrm{d}\xi\, \left(a_{1/4}(\xi)+a_{3/4}(\xi)\right)\tau_0(\xi)\\
             &= \frac{13}{3} - \int_0^{\infty} \mathrm{d}\omega\,
                  \left(\mathrm{e}^{-|\omega|/4}+\mathrm{e}^{-3|\omega|/4}\right) \tau_0(\omega) = 2\pi-5\,.
\end{aligned}
\end{equation}
which coincides with (\ref{finf2}) for $z=3$
(the same result has recently been obtained using a slightly different method in \cite{ZSXYC22}).

\subsection{$O(5)$}
The Bethe equations for the $O(5)$ model read
\begin{equation}
    \label{bae50}
    \begin{aligned}
      & \left[f_{1/2}(\lambda^{(1)}_j)\right]^{2L}= \prod_{k\ne j}^{L-n_1} f_1(\lambda^{(1)}_j-\lambda^{(1)}_k) f_1(\lambda^{(1)}_j+\lambda^{(1)}_k)
        \prod_{k=1}^{L-n_1-n_2} f_{-1}(\lambda^{(1)}_j-\lambda^{(2)}_k) f_{-1}(\lambda^{(1)}_j+\lambda^{(2)}_k)\,,\\
        &\qquad\qquad j=1,\dots,L-n_1\,, \\
      & \prod_{k\ne j}^{L-n_1-n_2} f_{1/2}(\lambda^{(2)}_j-\lambda^{(2)}_k) f_{1/2}(\lambda^{(2)}_j+\lambda^{(2)}_k) 
         =\prod_{k=1}^{L-n_1} f_{1/2}(\lambda^{(2)}_j-\lambda^{(1)}_k) f_{1/2}(\lambda^{(2)}_j+\lambda^{(1)}_k)\,,\\
      &\qquad\qquad j=1,\dots,L-n_1-n_2\,.
    \end{aligned}
\end{equation}
The ground state for even length spin chains is in the sector $(n_1,n_2)=(0,0)$ with $\lambda^{(1)}_j\in \mathbb{R}^+$ and $\lambda^{(2)}_j$ arranged in 2-strings $\lambda^{(2)}_j\simeq \xi_j\pm i/4$, $\xi_j>0$.
The corresponding energy is again given by (\ref{ene-o22m}).
Following the same steps as for the $O(3)$ model above we obtain Bethe equations for the string coordinates from (\ref{bae50})
\begin{equation}
    \begin{aligned}
      &2L\phi_{1/2}(\lambda^{(1)}_j) = 2\pi \mathrm{Q}_j - \phi_{1/2}(\lambda^{(1)}_j)
        + \sum_{k=1}^L\left(\phi_1(\lambda^{(1)}_j-\lambda^{(1)}_k)+\phi_1(\lambda^{(1)}_j+\lambda^{(1)}_k)\right)\\
      &\qquad - \sum_{k=1}^{L/2}\left[ \phi_{3/4}(\lambda^{(1)}_j-\xi_k) + \phi_{3/4}(\lambda^{(1)}_j+\xi_k)
                                      +\phi_{1/4}(\lambda^{(1)}_j-\xi_k) + \phi_{1/4}(\lambda^{(1)}_j+\xi_k)\right]\,,
                                      \quad j=1,\dots,L\,,\\
      &\sum_{k=1}^{L} \left[ \phi_{3/4}(\xi_j-\lambda^{(1)}_k)+\phi_{3/4}(\xi_j+\lambda^{(1)}_k)
                            +\phi_{1/4}(\xi_j-\lambda^{(1)}_k)+\phi_{1/4}(\xi_j+\lambda^{(1)}_k) \right]\\
      &\qquad = 2\pi\overline{\mathrm{Q}}_j - \phi_{1/2}(\xi_j) 
        + \sum_{k=1}^{L/2}\left[ \phi_1(\xi_j-\xi_k)+\phi_1(\xi_j+\xi_k)
                                +2\left(\phi_{1/2}(\xi_j-\xi_k)+\phi_{1/2}(\xi_j+\xi_k)\right)
                                \right]\,,\\
      &\qquad\qquad j=1,\dots,L/2\,.
    \end{aligned}
\end{equation}
For $L\gg1$ the ground state densities $\rho_0(\lambda)$ of real roots from the first level Bethe equations and $\overline{\rho}_0(\xi)$ of two-strings from the second one are given in terms of the integral equations
\begin{equation}
    \begin{aligned}
      \rho_0(\lambda) =& \frac1\pi a_{1/2}(\lambda) 
      +\frac1{2\pi L}\left(a_1(\lambda)-a_{3/4}(\lambda)+a_{1/2}(\lambda)-a_{1/4}(\lambda)\right)\\
      &
      -\frac1{2\pi}\int_{-\infty}^\infty \mathrm{d}\lambda'\, a_1(\lambda-\lambda')\rho(\lambda')+\frac1{2\pi}\int_{-\infty}^\infty \mathrm{d}\xi'\,
        \left[a_{3/4}(\lambda-\xi')+a_{1/4}(\lambda-\xi')\right] \overline{\rho}_0(\xi)\,,\\
      \overline{\rho}_0(\xi) =&
      \frac1{2\pi L} \left(a_1(\xi)-a_{3/4}(\xi)+3a_{1/2}(\xi)-a_{1/4}(\xi)\right)
      +\frac1{2\pi}\int_{-\infty}^\infty \mathrm{d}\lambda'\, 
        \left[a_{3/4}(\xi-\lambda')+a_{1/4}(\xi-\lambda')\right] \rho_0(\lambda')\\
      &-\frac1{2\pi}\int_{-\infty}^\infty \mathrm{d}\xi'\,
        \left[2a_{1/2}(\xi-\xi')+a_{1}(\xi-\xi')\right] \overline{\rho}_0(\xi')\,.
    \end{aligned}
\end{equation}
Solving these equations the boundary contribution to the density $\rho_0$ of first level roots is found to be
\begin{equation}
    \tau_0(\omega) = -\mathrm{e}^{-|\omega|/8}\, \frac{\sinh(5|\omega|/8)-\cosh(\omega/8)}{\cosh(3\omega/4)}\,.
\end{equation}
Using this expression in (\ref{FREEosp22m}) yields the resulting surface energy of the $O(5)$ spin chain (and the $OSp(n|2m)$ superspin chains with $n-2m=5$)
\begin{equation}
    f_\infty = -1+\frac{2\sqrt{3}}{9}\pi + \frac{\pi}{3} + \frac13\left(\psi\left(\frac5{12}\right)-\psi\left(\frac{11}{12}\right)\right)\,
\end{equation}
in agreement with Eq.~(\ref{finf2}) for $z=5$.

%

\end{document}